\documentclass[preprint,5p]{elsarticle}
\usepackage{amsmath,amssymb}

\begin{document}
\begin{frontmatter}

\title{Yukawa Alignment in a Multi Higgs Doublet Model: An effective approach}
\author[rvt]{H.~Ser\^odio}
\ead{hserodio@cftp.ist.utl.pt}
\address[rvt]{Departamento de F\'{\i}sica and Centro de F\'{\i}sica Te\'{o}rica de Part\'{\i}culas, Instituto Superior T\'{e}cnico, Av. Rovisco Pais, 1049-001 Lisboa, Portugal}


\begin{abstract}
In the two Higgs doublet model, natural flavour conservation can be achieved through the use of a discrete $Z_2$ symmetry. A less restrictive condition is the requirement of alignment in the Yukawa sector. So far, alignment has been an anzatz, not rooted in a specific model. In this letter we present a model for alignment, which starts with $2+N$ Higgs doublets, with natural flavour conservation imposed by a discrete symmetry. Only two of these scalars couple to the fermions, the other $N$ scalars are in a hidden sector. Assuming that the two scalar doublets coupled to fermions are heavy, their decoupling leads to an effective Yukawa interaction. The latter connects the fermions and the scalars of the hidden sector, and exhibits the same Yukawa coupling matrix for each of the $N$ scalars.  
\end{abstract}

\end{frontmatter}

\section{Introduction}
The Standard Model (SM) of particle physics is very successful phenomenologically. However, there is a general consensus that it should be an effective theory because it does not explain basic issues such as neutrino masses, number of families and others. The SM imposes the existence of a scalar particle in nature, the Higgs, responsible for the breaking of the gauge group $SU(2)\times U(1)$ and for the masses of the gauge and fermion fields. Even though one scalar doublet is the minimal setup, the addition of extra Higgs fields is appealing from the theoretical and phenomenological point of view.  

One minimal extension is the Two Higgs Doublet Model (2HDM), which adds new neutral and charged scalar particles, allowing spontaneous CP violation, dark matter candidates, and many other interesting features. However, phenomena such as flavour changing neutral currents (FCNC), which are very suppressed in nature and appear only at loop level for both gauge and Higgs sector in the SM, have no natural suppression in these models without additional constraints. A simple way to obtain natural relations is through the use of symmetries in the Lagrangian; when they preclude FCNC it is said that the model has natural flavor conservation (NFC). Glashow and Weinberg~\cite{Glashow:1976nt}, and Paschos \cite{Paschos:1976ay}, pointed out that sequential extensions of the SM have a GIM-like mechanism~\cite{Glashow:1970gm} suppressing all direct neutral currents effects. According to these works, the criterion for NFC can be reduced to the statement that all the Yukawa matrices, $\mathbf{Y}_q^h$, are simultaneously diagonalizable in each charged sector, independently of the parameters of the model, i.e.
\begin{equation}
 \mathbf{U}_L^{q\dagger}\mathbf{Y}_q^h\mathbf{U}_R^{q}=\text{diagonal for all }h\,,
\end{equation}
where $q$ stands for the quarks and $h$ for the different Higgs fields, $\mathbf{U}_{L,R}^{q}$ are unitary matrices. This NFC condition could be easily implemented in a 2HDM through the use of a $Z_2$ symmetry, which imposes that only one scalar doublet couples to each of the two (up or down) quark types. Later, Gatto et al.~\cite{Gatto:1979mr} have shown that the use of discrete symmetries that allow the simultaneous diagonalization of the Yukawa coupling matrices, in a general case where the Higgs couples to both up- and down-type quarks, gives trivial mixing schemes or very large mixing angles. However, none of these solutions is the desired one, if one expects the mixing angles to be generated at tree level with only small radiative corrections coming from loops. 

Recently, Pich and Tuz\'on~\cite{Pich:2009sp} have proposed an anzatz where the two Yukawa matrices are proportional. This alignment leads to NFC and would allow, in principle, for the correct mixing scheme. Nevertheless, this anzatz is not stable under the renormalization group equations (RGE)~\cite{Ferreira:2010xe}. Thus, if we insist on implementing this anzatz, we need to extend the particle content. For example, in the minimal supersymmetric extension of the SM (MSSM) the scalar sector has two Higgs doublets. However, if one only knew the scalar sector, and had no information about the supersymmetric particles, we would get a theory that is not invariant under the RGEs \cite{Ferreira:2010jy}. Therefore, it is natural to think that, if the alignment condition is true, additional particles must be present so that the alignment obtained in this way is stable under the RGEs.  

In this letter we adopt an effective field theory approach to the problem of the Yukawa alignment in a multi Higgs doublet model (MHDM). We start by  writing an UV complete model with $2+N$ Higgs doublets and then decouple two of them, the heavy ones. The resulting total Lagrangian can be viewed as an expansion in $1/\Lambda$:
\begin{equation}
\mathcal{L}_{eff}=\mathcal{L}_0+\frac{1}{\Lambda}\mathcal{O}_{d=5}+\frac{1}{\Lambda^2}\mathcal{O}_{d=6}+...
\end{equation}  
with $\Lambda$ the scale of the heavy modes. The Lagrangian $\mathcal{L}_0$ contains the renormalizable terms, while the other terms have contributions from operators $\mathcal{O}$ with higher dimensions. For dimension five operators, there is only one possibility. However, for dimension six, the number increases drastically. Nevertheless, since we specify the UV model, the number of dimension six operators appearing in the effective Lagrangian will be considerably reduced. 

This letter is organized as follows. In Section 2 we present a renormalizable model based on a discrete symmetry and with $2+N$ Higgs doublets plus two real scalar singlets. The way this symmetry is implemented is very similar to what is done in the usual 2HDM with the $Z_2$ symmetry. Therefore, we follow the standard distinction in model types. We next show how the decoupling of the heavy states affects the low energy theory, and we analyze the possible way to break the discrete symmetry group. In Section 3 we describe the effective MHDM for the different model types. We show that only Type-II and Y models allow for the alignment of the Yukawa matrices and study some phenomenological constraints on the scale of the heavy states. At the end of this section, the particular case of 2HDM is studied in more detail. Conclusions are presented in Section 4.

\section{From a renormalizable to an effective model}
Our goal is to obtain at low energies an effective MHDM with the Yukawa coupling matrices aligned. For that, we will start from a model with $2+N$ Higgs doublets. Two of these scalar doublets will couple to the fermions, as in the usual 2HDM with a $Z_2$ symmetry, in order to forbid tree-level FCNC. The additional $N$ doublets have no coupling to the fermion fields and reside in a hidden sector where only interactions with the gauge and scalar fields are allowed. In order to have this hidden sector an additional $Z_2$ is needed. We shall also add to the particle content two additional real scalar singlets which will allow the connection between the heavy states and the $N$ scalar fields.   
We then construct the renormalizable model within this framework, and study the process of decoupling of two heavy states. Finally, we shall analyze the vacuum alignment of the real scalar singlet fields.

\subsection{Constructing the model}
The two scalar doublets that are coupled to the fermion sector will be denoted by $\Phi_i$, $i=1,2$; the additional $N$ scalar doublets by $\phi_a$, with $a=1,...,N$, and the two real scalar singlets by $\eta_1$ and $\eta_2$ (cf. Table~\ref{content}). The particle content for the fermion and gauge sector is exactly the same as in the SM. All fermion fields are three dimensional vectors in the flavour space. The left-handed fields are invariant under the group $Z_2\times Z_2$; the transformations of the right-handed ones are summarized in Table~\ref{type}.

\begin{table}[!h]
\begin{center}
{\small
\begin{tabular}{r||c|c||c|c|c|}
&$\Phi_1$&$\Phi_2$&$\phi_a$&$\eta_1$&$\eta_2$\\
\hline\hline
$SU(2)$&2&2&2&1&1\\
\hline
$U(1)$&1/2&1/2&1/2&0&0\\
\hline
$Z_2\times Z_2$&$(1,1)$&$(1,-1)$&$(-1,1)$&$(-1,1)$&$(-1,-1)$\\
\hline
\end{tabular}}
\end{center}
\caption{\label{content} Scalar content of the usual 2HDM plus hidden sector and their transformation under $Z_2\times Z_2$.}
\end{table}

\begin{table}[!h]
\begin{center}
\begin{tabular}{l||c|c|c|}
&$u_R$&$d_R$&$e_R$\\
\hline\hline
Type-I&$(1,1)$&$(1,1)$&$(1,-1)$\\
Type-II&$(1,-1)$&$(1,1)$&$(1,1)$\\
Type-X&$(1,-1)$&$(1,-1)$&$(1,1)$\\
Type-Y&$(1,-1)$&$(1,1)$&$(1,-1)$\\
Inert&$(1,1)$&$(1,1)$&$(1,1)$\\
\hline
\end{tabular}
\end{center}
\caption{\label{type} The five possible implementations of the $Z_2\times Z_2$ in the Yukawa sector.}
\end{table}

This model is very similar to the 2HDM in the Yukawa sector. This is one of the key aspects which will allow the alignment. It is now important to write the new relevant Lagrangian interactions. The relevant terms will be the scalar kinetic sector, the Yukawa interactions and the scalar potential. We start with the kinetic Lagrangian for the scalar sector. It is given by
\begin{align}
\begin{split}\label{kinet}
\mathcal{L}_k=&\left(D^\mu\Phi_i\right)^\dagger\left(D_\mu\Phi_i\right)+
\left(D^\mu\phi_a\right)^\dagger\left(D_\mu\phi_a\right)\\
&+\frac{1}{2}\left(\partial^\mu\eta_i\right)\left(\partial_\mu\eta_i\right)\,,
\end{split}
\end{align}
where $D_\mu$ is the usual SM covariant derivative. The sum over repeated indexes is always implicit in this letter. The scalar potential is blind to the type of model. We get 
\begin{align}\label{pot}
\nonumber V=&M_{i}^2 \left(\Phi_i^\dagger \Phi_i\right)+\lambda^\Phi_{ij,kl} \left(\Phi_i^\dagger \Phi_j\right)\left(\Phi_k^\dagger \Phi_l\right)+m_{ab}^2 \left(\phi_a^\dagger \phi_b\right)\\
\nonumber&+\lambda^\Phi_{ab,cd}\left(\phi_a^\dagger \phi_b\right)\left(\phi_c^\dagger \phi_d\right)+\mu_{1\eta}^2\,\eta_1^2+\mu_{2\eta}^2\,\eta_2^2
+\lambda^\eta_{1}\eta_1^4\\
\nonumber&+\lambda^\eta_{2}\eta_2^4+\lambda^{\prime\eta}\eta_1^2\eta_2^2+\left(\Phi_i^\dagger \Phi_i\right)\left(\lambda_i^{\Phi\eta_1}\eta_1^2+\lambda_i^{\Phi\eta_2}\eta_2^2\right)\\
\nonumber&+\lambda^{\Phi\phi}_{iab}\left(\Phi_i^\dagger \Phi_i\right)\left(\phi_a^\dagger \phi_b\right)+\left(\phi_a^\dagger \phi_b\right)\left(\lambda^{\phi\eta}_{ab}\eta_1^2+\lambda^{\prime\phi\eta}_{ab}\eta_2^2\right)\\
&-\mu_{i,a}\,\Phi_i^\dagger\phi_a\eta_i-\mu_{i,a}^\ast\,\phi^\dagger_a\eta_i\Phi_i\,,
\end{align}
with $i,j,k,l=1,2$ and $a,b,c,d=1,...,N$. The Yukawa Lagrangian will be different for each model type. The general form is
\begin{align}\label{lagY}
\begin{split}
-\mathcal{L}_Y=&\overline{Q^0_L}\,\mathbf{Y}_d\,\Phi_i\,d_R^0+\overline{Q^0_L}\,\mathbf{Y}_u\,\tilde{\Phi}_j\,u_R^0+\overline{\ell^0_L}\,\mathbf{Y}_\ell\,\Phi_k\,e_R^0\\
&+\text{h.c.}\,,
\end{split}
\end{align}
with $\tilde{\Phi}=i\sigma_2 \left(\Phi^\dagger\right)^T$. The Yukawa matrices $\mathbf{Y}_{u,d,\ell}$ are $n_g\times n_g$ general complex matrices in the flavour space, with $n_g$ the number of generations. The way $\Phi_1$ and $\Phi_2$ couple to the fermions in Eq.~\eqref{lagY} is 

\begin{align}
\begin{split}
\text{Type-I:}&\quad i,j=1\,,\,k=2\quad\text{Type-II:}\quad i,k=1\,,\,j=2\\
\text{Type-X:}&\quad i,j=2\,,\,k=1\quad\text{Type-Y:}\quad i=1\,,\,j,k=2\\
\text{Inert:}&\quad i,j,k=1
\end{split}
\end{align}

As already noted this UV complete model only allows two of the $2+N$ scalar doublets ($\Phi_1$ and $\Phi_2$) to interact with fermions. However, non-local interactions present at the UV scale can become effectively local at some IR level. We next explore this mechanism in more detail. Nevertheless, we already know that the effective model will not have tree-level FCNC since these are absent from the full model.  

\subsection{Integrating out $\Phi_1$ and $\Phi_2$}
The effects of heavy particles at low energy are well described by the decoupling process. The mass of the heavy particle becomes the cutoff of the theory and its low-energy effects are present in effective higher dimensional operators or corrections to renormalizable interactions.

In this section we assume $\Phi_1$ and $\Phi_2$, the fields that couple to the fermion sector, to be heavy so that they decouple from the theory. Their classical equations of motion can be obtained from Eqs.~\eqref{kinet}-\eqref{lagY}. For $\Phi_i$ one gets
\begin{align}
\nonumber&\left(D^\mu D_\mu+M_{i}^2+\lambda_{iab}^{\Phi\phi}\left(\phi_a^\dagger \phi_b\right)+\lambda_i^{\Phi\eta_1}\eta_1^2+\lambda_i^{\Phi\eta_2}\eta_2^2\right)\Phi_i=\\
&-\left(X_i-\mu_{i,a}\eta_i\phi_a+\lambda_{ij,kl}^\Phi\left(\Phi_k^\dagger\Phi_l\right)\Phi_j\right)\,,
\end{align}
where the Yukawa terms $X_i$ depend on the model type:
\begin{align}\label{Xtype}
\begin{split}
\text{Type-I:}&\quad 
\left\{
\begin{array}{l}
X_1=\overline{d^0_R}\,\mathbf{Y}_d^\dagger\,Q^0_L+\left(\overline{Q^0_L}\,\mathbf{Y}_u\epsilon\,u^0_R\right)^T\\
X_2=\overline{e^0_R}\,\mathbf{Y}_\ell^\dagger\,\ell^0_L
\end{array}\right.\\
\text{Type-II:}&\quad 
\left\{
\begin{array}{l}
X_1=\overline{d^0_R}\,\mathbf{Y}_d^\dagger\,Q^0_L+\overline{e^0_R}\,\mathbf{Y}_\ell^\dagger\,\ell^0_L\\
X_2=\left(\overline{Q^0_L}\,\mathbf{Y}_u\epsilon\,u^0_R\right)^T
\end{array}\right.\\
\text{Type-X:}&\quad 
\left\{
\begin{array}{l}
X_1=\overline{e^0_R}\,\mathbf{Y}_\ell^\dagger\,\ell^0_L\\
X_2=\overline{d^0_R}\,\mathbf{Y}_d^\dagger\,Q^0_L+\left(\overline{Q^0_L}\,\mathbf{Y}_u\epsilon\,u^0_R\right)^T
\end{array}\right.\\
\text{Type-Y:}&\quad 
\left\{
\begin{array}{l}
X_1=\overline{d^0_R}\,\mathbf{Y}_d^\dagger\,Q^0_L\\
X_2=\left(\overline{Q^0_L}\,\mathbf{Y}_u\epsilon\,u^0_R\right)^T+\overline{e^0_R}\,\mathbf{Y}_\ell^\dagger\,\ell^0_L
\end{array}\right.\\
\text{Inert:}&\quad 
\left\{
\begin{array}{l}
X_1=\overline{d^0_R}\,\mathbf{Y}_d^\dagger\,Q^0_L+\left(\overline{Q^0_L}\,\mathbf{Y}_u\epsilon\,u^0_R\right)^T\\
\quad\quad+\overline{e^0_R}\,\mathbf{Y}_\ell^\dagger\,\ell^0_L\\
X_2=0
\end{array}\right.
\end{split}
\end{align}

The classical equations of motion can be solved iteratively in a perturbative way by expanding the solution in powers of $1/M_i$. In leading order we get the classical solutions
\begin{equation}\label{psi}
\Phi_i^c\simeq\frac{-1}{M_{i}^2}\left[X_i-\mu_{i,a}\eta_i\phi_a\right]\,,\\
\end{equation}
where the parameters $\mu_{i,a}$ have dimension of mass. A convenient way to parametrize them is
\begin{equation}
\mu_{1,a}=\lambda_a M_1\,,\quad\mu_{2,a}=\lambda_a^\prime M_2\,\,,
\end{equation}
with the parameters $\lambda_a$ and $\lambda_a^\prime$ dimensionless.
 
We now use these solutions to integrate out the heavy fields $\Phi_i$ from the model, in leading order (tree level). We see that the interaction terms between the fermions and the scalars are canceled by the mass term in the effective potential. Therefore, the Yukawa interactions in this effective model are given by the linear terms $\Phi^\dagger _i\phi_a\eta_i$ in the full potential. The effective potential at leading order can now be written as
\begin{align}
\begin{split}
 V^{eff}\simeq&\left[m_{ab}^2+\left(\lambda^{\phi\eta}_{ab}-\lambda_a^\ast\lambda_b\right)\eta_1^2\right.\\
 &\left. +\left(\lambda^{\prime\phi\eta}_{ab}-\lambda_a^{\prime\ast}\lambda^\prime_b\right)\eta_2^2\right] \left(\phi_a^\dagger \phi_b\right)\\
&+\lambda^\Phi_{ab,cd}\left(\phi_a^\dagger \phi_b\right)\left(\phi_c^\dagger \phi_d\right)\\
&+\mu_{1\eta}^2\,\eta_1^2+\mu_{2\eta}^2\,\eta_2^2
+\lambda^\eta_{1}\eta_1^4+\lambda^\eta_{2}\eta_2^4+\lambda^{\prime\eta}\eta_1^2\eta_2^2\,,
\end{split}
\end{align}
while the effective fermion-scalar interaction Lagrangian becomes
\begin{equation}\label{lagraint}
-\mathcal{L}_{int}^{eff}\simeq\frac{\eta_1}{M_{1}}X_1^\dagger\sum_{a=1}^N\lambda_a\phi_a+\frac{\eta_2}{M_{2}}\sum_{a=1}^N\lambda_a^{\prime\ast}\phi^\dagger_a\,X_2+\text{h.c.}\,.
\end{equation} 
Due to the decoupling, new four fermion interactions appear,
\begin{equation}\label{4fermion}
\mathcal{L}_{int}^{4f}\simeq\frac{1}{M_{1}^2}
X_1^\dagger X_1+\frac{1}{M_{2}^2}X_2^\dagger X_2\,.
\end{equation}
All effective terms are $Z_2\times Z_2$ preserving. Furthermore, when the fields $\eta_i$ acquire a non-vanishing vacuum expectation value (vev),
the Lagrangian terms in Eq.~\eqref{lagraint} are analogous to Yukawa-type interactions.

\subsection{The breaking of $Z_2\times Z_2$}
The $\eta_i$ fields are singlets under the SM gauge group. Therefore, the requirement of a non-vanishing vev will only lead to the breaking of the $Z_2\times Z_2$, leaving the gauge sector invariant. In this section, we study the possibility that the $\eta_i$ fields acquire a vev before the $N$ scalar doublets break the gauge symmetry. At this stage $\left<\eta_i\right>=u_i$, $\left<\phi_a\right>=0$, and the relevant terms of the effective potential are the self-interacting terms of $\eta_i$, 
\begin{align}
V(\eta)=&m_{1\eta}^2\,\eta_1^2+m_{2\eta}^2\,\eta_2^2
+\lambda^\eta_{1}\eta_1^4+\lambda^\eta_{2}\eta_2^4+\lambda^{\prime\eta}\eta_1^2\eta_2^2\,.
\end{align}
The minimization equations $\partial V/\partial \eta_i=0$ lead at the vacuum point
\begin{align}
\begin{split}
&2m_{1\eta}^2\,u_1
+4\lambda_1^\eta\,u_1^3+2\lambda^{\prime\eta}u_1u_2^2=0\,,\\
&2m_{2\eta}^2\,u_2
+4\lambda_2^\eta\,u_2^3+2\lambda^{\prime\eta}u_2u_1^2=0\,.
\end{split}
\end{align}
In order to solve them one can parametrize $u_1$ and $u_2$ as
\begin{equation}
u_1=u\sin\alpha\quad \text{and}\quad u_2=u\cos\alpha\,.
\end{equation}
The solution is then given by
\begin{align}
\begin{split}
u^2=&\frac{-m_{1\eta}^2}{\lambda^{\prime\eta}+\left(2\lambda_1^\eta-\lambda^{\prime\eta}\right)\sin^2\alpha}\,,\\
\sin^2\alpha=&\frac{m_2^2\lambda^{\prime\eta}-2\lambda_2^\eta m_1^2}{\lambda^{\prime\eta}\left(m_1^2+m_2^2\right)-2\left(\lambda_2^\eta m_1^2+m_2^2\lambda_1^\eta\right)}\,.
\end{split}
\end{align}
In order to ensure that the solution of the previous system is a minimum of the potential we need to evaluate the second derivatives. We obtain the following mass matrix:
\begin{align}
\begin{pmatrix}
2m_{1\eta}^2+12\lambda_1^\eta u_1^2+2\lambda^{\prime\eta}u_2^2&4\lambda^{\prime\eta} u_1u_2\\
4\lambda^{\prime\eta} u_1u_2&2m_{2\eta}^2+12\lambda_2^\eta u_2^2+2\lambda^{\prime\eta}u_1^2
\end{pmatrix}\,.
\end{align}
As a simple exercise, one can look for a region of parameters where the non-trivial solution is favored.  For example, in the region of parameters with $\lambda^{\prime\eta}<<1$, the two scalars decouple from each other and we get a non-trivial solution
\begin{equation}
u_i\simeq\sqrt{-\frac{m^2_{i\eta}}{2\lambda^\eta_i}}\,,
\end{equation}
with $m_{i\eta}^2<0$. The mass spectrum becomes $m_{i}^2\simeq-4m_{i\eta}^2$, and leads to a ground state value of the potential
\begin{align}\label{vevu}
<V>=&-\frac{m_{1\eta}^4}{4\lambda^\eta_1}-\frac{m_{2\eta}^4}{4\lambda^\eta_2}\,.
\end{align}
From this simple analysis it becomes clear that in certain parameter regions, the potential $V(\eta)$ is minimized by non-vanishing vevs.

\section{The effective Multi Higgs Doublet Model}
\subsection{Effective MHDM Lagrangian}
When $\eta_i$ breaks the discrete symmetry by acquiring a vev, the scalar potential for the $N$ Higgs doublets become
\begin{align}\label{Veff}
V^{eff}\simeq&\left(M_\phi^2\right)_{ab}\left(\phi_a^\dagger \phi_b\right)+\lambda^\phi_{ab,cd} \left(\phi_a^\dagger \phi_b\right)\left(\phi_c^\dagger \phi_d\right)\,,
\end{align}
with
\begin{align}
\begin{split}
\left(M_\phi^2\right)_{ab}=&m_{ab}^2+\left(\lambda^{\phi\eta}_{ab}-\lambda_a^\ast\lambda_b\right)u_1^2+\left(\lambda^{\prime\phi\eta}_{ab}-\lambda_a^{\prime\ast}\lambda^\prime_b\right)u_2^2\,.
\end{split}
\end{align}
The most important changes, which are due to the decoupling of $\Phi_i$ and the breaking of discrete symmetry $Z_2\times Z_2$, are present in the new effective Yukawa Lagrangian
\begin{equation}\label{effYlagra}
-\mathcal{L}_{Y}^{eff}\simeq\frac{u_1}{M_{1}}X_1^\dagger\sum_{a=1}^N\lambda_a\phi_a+\frac{u_2}{M_{2}}\sum_{a=1}^N\lambda_a^{\prime\ast}\phi^\dagger_a\,X_2+\text{h.c}\,.
\end{equation} 
Depending on the implementation of the $Z_2\times Z_2$ symmetry in the UV model the interaction in the fermionic-scalar sector at low energy will have different consequences. 

According to Eq.~\eqref{effYlagra}, one could naively expect a natural alignment of the Yukawa matrices in all the different implementation types. Note, however, that due to the freedom we have to redefine the $N$ scalar doublets by a unitary transformation, this is not completely true. The integration of the heavy fields splits the five types of models into three different sets:
\begin{itemize}
\item \textit{Inert quark sector}: The Type-I and Type-X can be put into a basis where only the combination $\phi^\prime\propto\sum_a\lambda_a\phi_a$ or $\phi^\prime\propto\sum_a\lambda^\prime_a\phi_a$ couples in the quark sector. All the scalar fields couple to the leptons.
\item \textit{Inert lepton sector}: The Type-II and Type-Y can be put in a basis where the leptons couple to only one scalar combination, but one of the quark sectors, up or down, couple in an aligned way to the $N$ Higgs fields.
\item \textit{Inert}: The Inert model can always be put in a basis where all the fermions couple to only one scalar combination. Therefore, the UV Inert model stays Inert after the decoupling. 
\end{itemize}
\begin{table}[h]
\begin{center}
\begin{tabular}{l||c|c|c|}
&$c_u$&$c_d$&$c_\ell$\\
\hline\hline
Type-I$\begin{array}{c}\\\\\end{array}$&$e^{-i\theta_a}\lambda^\ast_av_a$&$e^{i\theta_a}\lambda_av_a$&$e^{i\theta_a}\lambda^\prime_av_a$\\
Type-II$\begin{array}{c}\\\\\end{array}$&$e^{-i\theta_a}\lambda_a^{\prime\ast} v_a$&$e^{i\theta_a}\lambda_av_a$&$e^{i\theta_a}\lambda_av_a$\\
Type-X$\begin{array}{c}\\\\\end{array}$&$e^{-i\theta_a}\lambda_a^{\prime\ast} v_a$&$e^{i\theta_a}\lambda^\prime_av_a$&$e^{i\theta_a}\lambda_av_a$\\
Type-Y$\begin{array}{c}\\\\\end{array}$&$e^{-i\theta_a}\lambda_a^{\prime\ast} v_a$&$e^{i\theta_a}\lambda_av_a$&$e^{i\theta_a}\lambda^\prime_av_a$\\
Inert$\begin{array}{c}\\\\\end{array}$&$e^{-i\theta_a}\lambda_a^\ast v_a$&$e^{i\theta_a}\lambda_av_a$&$e^{i\theta_a}\lambda_av_a$\\
\hline
\end{tabular}
\caption{\label{cparameters} Parameters $c$'s for each of the types of models. }\end{center}
\end{table}
After the spontaneous breaking of the electroweak symmetry, i.e. $\left<\phi_a\right>=(0,v_ae^{i\theta_a}/\sqrt{2})$, one gets the following mass matrices
\begin{equation}\label{massmatrix}
\mathbf{M}_f=\frac{c_f}{\sqrt{2}}\frac{u_i}{M_i}\mathbf{Y}_f\,,
\end{equation}
with $f=\left\{u,d,\ell\right\}$, and $i=1$ or $2$ if the parameter  $c_f$ has the constant coupling $\lambda$ or $\lambda^\prime$, respectively. The parameters $c_f$ depend on the type of model implemented. Their explicit forms are presented in Table~\ref{cparameters}.   In general, the mass matrices are bi-diagonalizable as
\begin{equation}\label{diagM}
\mathbf{U}_L^{f\dagger}\mathbf{M}_f\mathbf{U}_R^f=\mathbf{D}_f\,,
\end{equation}
with $\mathbf{D}_f$ a real diagonal matrix containing the masses of the fermion fields. This bi-diagonalization can be achieved through the field transformations 
\begin{align}\label{massbasis}
\begin{split}
&u_{L,R}^0\rightarrow \mathbf{U}^u_{L,R} u_{L,R}\,,\quad d_{L,R}^0\rightarrow \mathbf{U}^u_{L,R} d_{L,R}\\
&e_{L,R}^0\rightarrow\mathbf{U}_{L,R}^\ell e_{L,R}\,.
\end{split}
\end{align} 
When we perform these transformations in the effective model, all the neutral scalar components of $\phi_a$ have the couplings to the fermions diagonal. As an example, the particular case of two Higgs doublets will be given at the end of this section.

Finally, a few words should be addressed to the scale at which the discrete symmetry is broken. We notice that the mass matrix in Eq.~\eqref{massmatrix} is the leading source contributing to the fermion masses in this model. Therefore, in order to guarantee the perturbative regime for the Yukawa couplings one needs
\begin{equation}
\mathcal{O}\left(|c_f|\frac{u_i}{M_i}\right)\sim v\,,
\end{equation}
with $v=\left(\sum_av_a^2\right)^{1/2}=246\,\text{GeV}$. This requirement can be easily satisfied when $\mathcal{O}\left(u_i\right)\lesssim\mathcal{O}\left(M_i\right)$ and the dimensionless parameters $\lambda$ and $\lambda^\prime$ are of order one. Due to the freedom in the model, the previous requirements have no implications in the way the heavy scalars in Eq.~\eqref{psi} have been integrated out. Lower bounds for the heavy scalar fields can be obtained from the four fermion interactions.

\subsection{Bounds for the heavy scale}
The four fermion interactions in Eq.~\eqref{4fermion} have a very fundamental difference from the usual dimension six operators \cite{Buchmuller:1985jz}. In this model, they are proportional to the Yukawa matrices [see Eq.~\eqref{Xtype}]. Going to the mass basis through the field transformations in Eq.~\eqref{massbasis}, the interactions become proportional to $y_\alpha y_\beta$, where $y_\alpha$ are the Yukawa couplings of the fermions, approximately given by
\begin{align}\label{yukawas}
\begin{split}
&\mathcal{O}\left(y_e\right)\sim\mathcal{O}\left(y_u\right)\sim\mathcal{O}\left(y_d\right)\sim10^{-5}\,,\\
&\mathcal{O}\left(y_\mu\right)\sim\mathcal{O}\left(y_s\right)\sim 10^{-3}\,,\\
&\mathcal{O}\left(y_\tau\right)\sim\mathcal{O}\left(y_c\right)\sim\mathcal{O}\left(y_b\right)\sim10^{-2}\,,\\
&\mathcal{O}\left(y_t\right)\sim1\,.
\end{split}
\end{align}
These interaction terms are of the form $\left(\overline{\Phi_L}\chi_R\right)\left(\overline{\chi_R}\Phi_L\right)$, with $\Phi_L$ a left-handed doublet of $SU(2)$ and $\Phi_R$ a right-handed singlet. Using a Fierz rearrangement, we are able to transform this term in 
\begin{equation}
-\frac{1}{2}\left[\left(\overline{\Phi_{1}}\gamma^\mu P_L\Phi_1\right)+\left(\overline{\Phi_{2}}\gamma^\mu P_L\Phi_2\right)\right]\left(\overline{\chi}\gamma_\mu P_R\chi\right)\,.
\end{equation}
The possible four fermion interactions, for each Type of model, are presented in Table~\ref{fourfermionint}.
\begin{table}[h]
\begin{tabular}{r||c|c|c|c|c|}
&I&II&X&Y&Inert\\
\hline\hline
&&&&& \\[-2ex]
$\left(\overline{u_\alpha}\gamma^\mu P_L u_\beta\right)\left(\overline{d_\beta}\gamma_\mu P_R d_\alpha\right)$&$\surd$&$\surd$&$\surd$&$\surd$&$\surd$\\
$\left(\overline{d_\alpha}\gamma^\mu P_Ld_\beta\right)\left(\overline{d_\beta}\gamma_\mu P_R d_\alpha\right)$&$\surd$&$\surd$&$\surd$&$\surd$&$\surd$\\
$\left(\overline{u_\alpha}\gamma^\mu P_L u_\beta\right)\left(\overline{u_\beta}\gamma_\mu P_Ru_\alpha\right)$&$\surd$&$\surd$&$\surd$&$\surd$&$\surd$\\
$\left(\overline{d_\alpha}\gamma^\mu P_L d_\beta\right)\left(\overline{u_\beta}\gamma_\mu P_Ru_\alpha\right)$&$\surd$&$\surd$&$\surd$&$\surd$&$\surd$\\
$\left(\overline{d_\alpha}P_L d_\alpha\right)\left(\overline{u_\beta}P_Lu_\beta\right)$&$\surd$&&$\surd$&&$\surd$\\
$\left(\overline{d_\alpha}P_L u_\alpha\right)\left(\overline{u_\beta}P_Ld_\beta\right)$&$\surd$&&$\surd$&&$\surd$\\
&&&&& \\[-2ex]
\hline
&&&&& \\[-2ex]
$\left(\overline{\nu_\alpha}\gamma^\mu P_L\nu_\beta\right)\left(\overline{e_\beta}\gamma_\mu P_R e_\alpha\right)$&$\surd$&$\surd$&$\surd$&$\surd$&$\surd$\\
$\left(\overline{e_\alpha}\gamma^\mu P_Le_\beta\right)\left(\overline{e_\beta}\gamma_\mu P_R e_\alpha\right)$&$\surd$&$\surd$&$\surd$&$\surd$&$\surd$\\
&&&&& \\[-2ex]
\hline
&&&&& \\[-2ex]
$\left(\overline{\nu_\alpha} P_Re_\alpha\right)\left(\overline{d_\beta}P_L u_\beta\right)$&&$\surd$&&&$\surd$\\
$\left(\overline{e_\alpha}P_Re_\alpha\right)\left(\overline{d_\beta}P_L d_\beta\right)$&&$\surd$&&&$\surd$\\
$\left(\overline{\nu_\alpha} P_Re_\alpha\right)\left(\overline{d_\beta}P_R u_\beta\right)$&&&&$\surd$&$\surd$\\
$\left(\overline{e_\alpha}P_Re_\alpha\right)\left(\overline{u_\beta}P_R u_\beta\right)$&&&&$\surd$&$\surd$\\
\hline
\end{tabular}
\caption{\label{fourfermionint} All four fermion operators present in this model.}
\end{table}
From decay processes of mesons and leptons, several bounds can be found \cite{Raidal:2008jk}. However, the particular dependence of these terms on the Yukawa couplings leads to a much less restrictive lower bound. For example, in the case of Type-II, the quantity $R_\pi\equiv\Gamma\left(\pi^+\rightarrow e^+\nu_e\right)/\Gamma \left(\pi^+\rightarrow \mu^+\nu_\mu\right)$ is one of the most restrictive. Following Ref.~\cite{Raidal:2008jk} we get from this observable the constraint
\begin{equation}
\frac{C_{qde}^{eedu}}{M_1^2}\leq\frac{4G_F}{\sqrt{2}}\epsilon_{qde}^{eedu}
\end{equation}
with $\epsilon_{qde}^{eedu}=1.5\times 10^{-7}$ and $G_F=1.16637\times 10^{-5}\, \text{GeV}^{-2}$ the Fermi constant. The lower bound that we get is
\begin{equation}
M_1\geq 450\,\sqrt{y_ey_d}\,\text{TeV}\sim 4.5 \, \text{GeV}\,,
\end{equation}
which is not restrictive at all.

\subsection{A particular case: Effective 2HDM}
We now turn our attention to the particular case of Type-II 2HDM. In this case, the scalar potential is given by Eq.~\eqref{Veff} and the Yukawa Lagrangian, in Eq.~\eqref{effYlagra}, can be written as
\begin{align}
\begin{split}
-\mathcal{L}_{\text{Yuk}}^{2HDM}=&\frac{u_1}{M_1}\overline{Q_L^0}\left(\mathbf{\Gamma}_1\phi_1+\mathbf{\Gamma}_2\phi_2\right)d_R^0\\
&+\frac{u_2}{M_2}\overline{Q_L^0}\left(\mathbf{\Delta}_1\tilde{\phi}_1+\mathbf{\Delta}_2\tilde{\phi}_2\right)u_R^0\\
&+\frac{u_1}{M_1}\overline{\ell_L^0}\left(\mathbf{\Pi}_1\phi_1+\mathbf{\Pi}_2\phi_2\right)e_R^0+\text{h.c.}\,,
\end{split}
\end{align}
where we have defined
\begin{align}\label{relations}
\mathbf{\Gamma}_a=\lambda_a\mathbf{Y}_d\,,\quad \mathbf{\Delta}_a=\lambda^{\prime\ast} _a\mathbf{Y}_u\quad\text{and}\quad \mathbf{\Pi}_a=\lambda_a\mathbf{Y}_\ell\,.
\end{align} 
The type-Y 2HDM can be similarly obtained through the change $\mathbf{\Pi}_a=\lambda^\prime_a\mathbf{Y}_\ell$ in Eq.~\eqref{relations}. It is convenient to change basis in the scalar sector and go to the so-called Higgs basis. We perform that through the orthogonal rotation  
\begin{equation}
\begin{pmatrix}
H_1\\
H_2
\end{pmatrix}=
\frac{1}{v}
\begin{pmatrix}
v_1&v_2\\
v_2&-v_1
\end{pmatrix}
\begin{pmatrix}
\phi_1\\
e^{-i\alpha}\phi_2
\end{pmatrix}\,,
\end{equation}
with the new fields parametrized as
\begin{equation}
H_1=\frac{1}{\sqrt{2}}
\begin{pmatrix}
\sqrt{2}\,G^+\\
v+H^0+iG^0
\end{pmatrix}\,,\quad
H_2=\frac{1}{\sqrt{2}}
\begin{pmatrix}
\sqrt{2}\,H^+\\
R+iI
\end{pmatrix}\,.
\end{equation}
In this new basis only the first field has a vev given by $v=\sqrt{v_1^2+v_2^2}$. The scalar components $G^+$ and $G^0$ are the Goldstone bosons and $\alpha=\theta_2-\theta_1$. The Yukawa Lagrangian can now be written in terms of the new components. We get
\begin{align}
\nonumber-\mathcal{L}_{\text{Yuk}}=&...+\frac{\sqrt{2}}{v}H^+
\left[\overline{u_L^0}\mathbf{N}_dd_R^0-\overline{u_R^0}\mathbf{N}^\dagger_ud_L^0
+\overline{\nu_L^0}\mathbf{N}_\ell e_R^0\right]\\
\nonumber&+\frac{R}{v}\left[\overline{d_L^0}\mathbf{N}_dd_R^0+\overline{u_L^0}\mathbf{N}_uu_R^0
+\overline{e_L^0}\mathbf{N}_\ell e_R^0\right]\\
\nonumber&+\frac{iI}{v}\left[\overline{d_L^0}\mathbf{N}_dd_R^0-\overline{u_L^0}\mathbf{N}_uu_R^0+\overline{e_L^0}\mathbf{N}_\ell e_R^0\right]\\
\nonumber&+\left(1+\frac{H^0}{v}\right)\left[\overline{d_L^0}\mathbf{M}_dd_R^0+\overline{u_L^0}\mathbf{M}_uu_R^0+\overline{e_L^0}\mathbf{M}_\ell e_R^0\right]\\
&+\text{h.c.}\,,
\end{align}
with the mass matrices given by
\begin{align}
\begin{split}
\mathbf{M}_u=&\frac{1}{\sqrt{2}}\frac{u_2}{M_2}\left(\mathbf{\Delta}_1 v_1+e^{-i\alpha}\mathbf{\Delta}_2 v_2\right)\,,\\
\mathbf{M}_{d}=&\frac{1}{\sqrt{2}}\frac{u_1}{M_1}\left(\mathbf{\Gamma}_1 v_1+e^{i\alpha}\mathbf{\Gamma}_2 v_2\right)\,,\\
\mathbf{M}_{\ell}=&\frac{1}{\sqrt{2}}\frac{u_1}{M_1}\left(\mathbf{\Pi}_1 v_1+e^{i\alpha}\mathbf{\Pi}_2 v_2\right)\,,
\end{split}
\end{align}
and the new flavour-changing matrices
\begin{align}
\begin{split}
\mathbf{N}_u=&\frac{1}{\sqrt{2}}\frac{u_2}{M_2}\left(\mathbf{\Delta}_1 v_2-e^{-i\alpha}\mathbf{\Delta}_2 v_1\right)\,,\\
\mathbf{N}_{d}=&\frac{1}{\sqrt{2}}\frac{u_1}{M_1}\left(\mathbf{\Gamma}_1 v_2-e^{i\alpha}\mathbf{\Gamma}_2 v_1\right)\,,\\
\mathbf{N}_{\ell}=&\frac{1}{\sqrt{2}}\frac{u_1}{M_1}\left(\mathbf{\Pi}_1 v_2-e^{i\alpha}\mathbf{\Pi}_2 v_1\right)\,.
\end{split}
\end{align}
In general, while the mass matrices are bidiagonalizable through Eq.~\eqref{diagM}, the matrices $\mathbf{N}_f$ are not. This fact usually leads to undesired FCNC. However, the way this model is implemented leads, through Eq.~\eqref{relations}, to the relations
\begin{align}\label{gammaprop}
\mathbf{\Gamma}_2=\xi_{d}\mathbf{\Gamma}_1\,,\quad \mathbf{\Delta}_2=\xi_u^\ast\mathbf{\Delta}_1\quad\text{and}\quad \mathbf{\Pi}_2=\xi_\ell\mathbf{\Pi}_1\,,
\end{align}
with
\begin{equation}
\xi_{d\,,\ell}=\frac{\lambda_2}{\lambda_1}\quad\text{and}\quad \xi_u=\frac{\lambda_2^\prime}{\lambda^\prime_1}\,.
\end{equation}
Through Eq.~\eqref{gammaprop}, we can relate the mass and the flavour-changing matrices 
\begin{equation}
\mathbf{N}_u=\varsigma_u^\ast \mathbf{M}_u\,,\quad \mathbf{N}_d=\varsigma_d \mathbf{M}_d\,,\quad \mathbf{N}_\ell=\varsigma_\ell \mathbf{M}_\ell\,,
\end{equation} 
with the proportionality factors given by
\begin{equation}\label{complexparam}
\varsigma_f=\frac{\tan\beta-e^{i\alpha}\xi_f}{1+e^{i\alpha}\xi_f \tan\beta}\,,\quad\quad\tan\beta=\frac{v_2}{v_1}\,.
\end{equation}
Only two of the usual three complex parameters $\varsigma_f$ are free in this model. Their phases provide new sources of CP violation without tree-level FCNCs.

Applying the field transformations in Eq.~\eqref{massbasis} we can go to the physical basis. The only flavour-changing structure is the CKM matrix present in the charged sector. The relevant Lagrangian is of the form
\begin{align}
\begin{split}
-\mathcal{L}_{\text{Yuk}}^{FC}=&\frac{\sqrt{2}}{v} H^+
\left[\varsigma_d\overline{u_L}\,\mathbf{V}\mathbf{D}_d\,d_R-\varsigma_u^\ast\overline{u_R}\,\mathbf{D}_u\mathbf{V}\,d_L
\right.\\
&\left.+\varsigma_\ell\overline{\nu_L}\,\mathbf{D}_\ell \,e_R\right]+\text{h.c.}\,,
\end{split}
\end{align}
with $\mathbf{V}=\mathbf{U}_L^{u^\dagger}\mathbf{U}_L^{d}$ the CKM matrix. This model represents an example of the aligned 2HDM introduced by Pich and Tuz\'on \cite{Pich:2009sp} and Jung, Pich and Tuz\'on \cite{Jung:2010}, where in the last reference a more detailed study of the phenomenological implications was done. Due to the particular form of the complex parameters $\varsigma_f$ in Eq.~\eqref{complexparam}, this model allows the new CP violating phases to be spontaneously generated when the electroweak symmetry is broken by $\phi_1$ and $\phi_2$. 

The possibility of having simultaneously alignment and spontaneous CP violation is one of the most interesting features of this type of approach. In general, whenever one imposes on the 2HDM Lagrangian a discrete symmetry together with the CP symmetry, the product of these two symmetries will also be another CP symmetry \cite{Branco:1999fs}. Therefore, in order for spontaneous CP violation to occur, the vacuum has to break both symmetries. Within 2HDM this is not possible because of the very restrictive form of the scalar potential. The approach taken here avoids this problem. The scalars that couple to the fermions at the UV level are not the ones that couple to them at low scales. This leaves the scalar sector of $\phi_1$ and $\phi_2$ the most general one, even though a discrete symmetry is imposed to get NFC. At the renormalizable level, imposing CP invariance under the trivial CP transformation corresponds to setting all coupling constants real. However, due to the richness of the scalar potential of Eq.~\eqref{Veff}, a non-trivial phase $\alpha$ can be in general obtained and CP be spontaneously broken.

\section{Conclusions}
In the context of a multi Higgs scenario, we have presented in this letter a simple model where all the effective Yukawa coupling matrices are aligned. The alignment is a consequence of natural flavour conservation at the UV scale, imposed by a discrete symmetry, and the decoupling of heavy states. Since only two heavy Higgs doublets couple to fermions, the way to implement the discrete symmetry is similar to what is done in the standard two Higgs doublet models. We have found that models starting from Type-I, Y and Inert implementations do not lead to an alignment. On the other hand, Type-II and X implementations do lead to an effective model with a Yukawa alignment. 

The particular case of the two Higgs doublet model in a Type-II framework was presented. As expected, flavour changing neutral currents are absent due to the relation $\mathbf{N}_f=\varsigma_f \mathbf{M}_f$ between the mass and flavour-coupling matrices. In this model the complex parameters $\varsigma_f$ are not all independent, as in the anzatz of Ref.~\cite{Pich:2009sp}. For the Type-II (Type-X) implementation we have $\varsigma_d=\varsigma_\ell$ ($\varsigma_u^\ast=\varsigma_\ell$). This effective two Higgs doublet model allows for spontaneous CP violation and natural flavour conservation.

In this letter we have presented a possible mechanism for Yukawa coupling alignment in a multi Higgs doublet scenario. This was done through the explicit construction of an effective aligned multi Higgs doublet model, taking into account only leading order effects. For a more rigorous study, we would have to go further in the expansion of the heavy fields and consider loop corrections. We note, however, that although new interactions will arise, the new contributions will not spoil the Yukawa coupling alignment. The possibility of having alignment to all orders is the most interesting feature of this model.

\section*{Acknowledgements}
I am very grateful to R. Gonz\'alez Felipe and to J.P. Silva for their invaluable discussions, support and careful revision of the manuscript. I am also grateful to L. Lavoura for useful comments during the initial stage of this work. 
This work was supported by  Funda\c{c}\~{a}o para a Ci\^{e}ncia e a Tecnologia (FCT, Portugal) under the Grant No. SFRH/BD/36994/2007 and, partially, through the projects PTDC/FIS/098188/2008, CFTP-FCT Unit 777 and CERN/FP/116328/2010, which are funded by POCTI (FEDER).


\begin{thebibliography}{10}
\bibitem{Glashow:1976nt}
  S.~L.~Glashow, S.~Weinberg,
  Phys.\ Rev.\  {\bf D15 } (1977)  1958.
  
\bibitem{Paschos:1976ay}
  E.~A.~Paschos,
  Phys.\ Rev.\  {\bf D15 } (1977)  1966.
  
\bibitem{Glashow:1970gm}
  S.~L.~Glashow, J.~Iliopoulos, L.~Maiani,
  Phys.\ Rev.\  {\bf D2 } (1970)  1285-1292.
  
\bibitem{Gatto:1979mr}
  R.~Gatto, G.~Morchio, G.~Sartori, F.~Strocchi,
  Nucl.\ Phys.\  {\bf B163 } (1980)  221.
  
\bibitem{Pich:2009sp}
  A.~Pich, P.~Tuz\'on,
  Phys.\ Rev.\  {\bf D80 } (2009)  091702.
  [arXiv:0908.1554 [hep-ph]].
  
  
\bibitem{Ferreira:2010xe}
  P.~M.~Ferreira, L.~Lavoura, J.~P.~Silva,
  Phys.\ Lett.\  {\bf B688 } (2010)  341-344.
  [arXiv:1001.2561 [hep-ph]].
  
\bibitem{Ferreira:2010jy}
  P.~M.~Ferreira, H.~E.~Haber, J.~P.~Silva,
  Phys.\ Rev.\  {\bf D82 } (2010)  016001.
  [arXiv:1004.3292 [hep-ph]].
  
\bibitem{Buchmuller:1985jz}
  W.~B\"uchmuller, D.~Wyler,
  Nucl.\ Phys.\  {\bf B268 } (1986)  621.
  
\bibitem{Raidal:2008jk}
  M.~Raidal, A.~van der Schaaf, I.~Bigi, M.~L.~Mangano, Y.~K.~Semertzidis, S.~Abel, S.~Albino, S.~Antusch {\it et al.},
  Eur.\ Phys.\ J.\  {\bf C57 } (2008)  13-182.
  [arXiv:0801.1826 [hep-ph]].
  
 \bibitem{Jung:2010}
  M.~Jung, A.~Pich, P.~Tuz\'on,
  JHEP {\bf 1011 } (2010)  003.
  [arXiv:1006.0470 [hep-ph]].
   
\bibitem{Branco:1999fs}
  G.~C.~Branco, L.~Lavoura, J.~P.~Silva,
  Int.\ Ser.\ Monogr.\ Phys.\  {\bf 103 } (1999)  1-536.
  
\end{thebibliography}
\end{document}